\documentclass[final,3p,times]{elsarticle}

\usepackage{amssymb}

\usepackage{amsmath}
\usepackage{xcolor}

\def\pr{{\partial}}

\newcommand\Real{\mbox{Re}}

\newcommand\Rey{\mbox{\textit{Re}}}

\journal{European Journal of Mechanics. B/Fluids }

\begin{document}

\begin{frontmatter}


\title{On the stability of the Couette-Taylor flow
between rotating porous cylinders with radial flow}
\author[York]{Konstantin Ilin\corref{cor1}}
\ead{konstantin.ilin@york.ac.uk}
\author[Rostov,Vladik]{Andrey Morgulis}
\ead{amor@math.sfedu.ru}
\cortext[cor1]{Corresponding author}
\address[York]{Department of Mathematics, University of York, Heslington, York YO10 5DD, UK}
\address[Rostov]{Department of Mathematics, Mechanics and Computer Science, The Southern Federal University, Rostov-on-Don, Russian Federation}
\address[Vladik]{South Mathematical Institute, Vladikavkaz Center of RAS, Vladikavkaz, Russian Federation}

\begin{abstract}
We study the stability of the Couette-Taylor flow between porous cylinders with radial throughflow.
It had been shown earlier that this flow can be unstable
with respect to non-axisymmetric (azimuthal or helical) waves provided that
the radial Reynolds number, $R$
(constructed using the radial velocity at the inner cylinder and
its radius), is high. In this paper, we present a very detailed and, in
many respects, novel chart of critical curves in a region of
moderate values of $R$, and we show that, starting from values of $R$, as low as $10$, the
critical modes inherited from the inviscid instability gradually
substitute the classical Taylor vortices. Also, we have looked
more closely at the effect of a weak radial flow (relatively low $R$) on the Taylor
instability and found that a radial flow directed from the inner
cylinder to the outer one is capable of stabilizing the Couette-Taylor flow provided that
the gap between the cylinders is wide enough. This observation is in a
sharp contrast with the case of relatively narrow gaps for which the opposite effect
is well-known.
\end{abstract}


\end{frontmatter}



\section{Introduction}
\label{sec:Intro}

We study the linear stability of a steady viscous incompressible flow in a gap between two rotating porous cylinders with a radial flow.
The governing parameters
of the flow are the radial Reynolds number, $R$ (constructed using the radial velocity at the inner cylinder and its radius), and two azimuthal
Reynolds numbers, $\Rey_1$ and $\Rey_2$ (based on the azimuthal velocities of the inner and outer cylinders, respectively, and the gap between the cylinders).
The velocity of the basic flow has nonzero azimuthal and radial components and zero axial component. It is a straightforward generalisation of
the classical Couette-Taylor flow to the case
when a radial flow is present. The direction of the radial flow can be from
the inner cylinder to the outer one (the diverging flow) or from the outer cylinder
to the inner one (the converging flow). The most interesting feature of the basic flow is that the
azimuthal velocity profile non-trivially depends on the radial Reynolds number $R$. In particular,
for high radial Reynolds numbers, it becomes irrotational everywhere except for
a thin boundary layer near the outflow part of the boundary (i.e. near the outer cylinder for the diverging flow or
near the inner cylinder for the converging flow).

The stability of the Couette-Taylor flow with a radial throughflow had been studied by many authors
(e.g., \cite{Bahl, Chang, Min, Kolyshkin, Kolesov, Serre, Martinand, Martinand2017}).
Most studies were motivated by applications to dynamic filtration devices (see, e.g., \cite{Wron'ski,Beadoin})
{and vortex flow reactors (see \cite{Giordano} and references therein).
Recently, it was also argued by \citet{Gallet2010} and \citet{Kerswell} that
such flows may have some relevance to astrophysical flows in accretion discs
(see also \cite{Kersale}). These possible applications make it important to
better understand the structure and the stability properties of the Couette-Taylor flow with radial flow
not only for flow regimes specific to a particular application or a device, but also for a much wider
range of the governing parameters of the problem.}

One of the main aims of early papers (e.g., \cite{Bahl, Min, Kolyshkin})
was to determine the effect of the radial flow
on the stability of the circular Couette-Taylor flow to axisymmetric perturbations, and the common conclusion was
that both a converging radial flow and a sufficiently strong diverging
flow have a stabilizing effect on the Taylor instability, but when a diverging flow is weak, it has a destabilizing effect
\cite{Min, Johnson, Kolyshkin}. {We note in passing that we have found that,
when the cylinders rotate in opposite directions,
the destabilising effect of a weak diverging flow
occurs for all values of the gap between the cylinders; however, when the outer cylinder is not rotating, a weak diverging
flow has a destabilising effect, but only if the gap between the cylinders is relatively small, and it has a stabilising
effect if the gap is sufficiently large (see section 3.2 of the present paper).}
The question that remained open was whether a radial flow itself can induce instability for flows
which are stable without it. This question had been answered affirmatively
by \citet{Fujita} and later by \citet{Gallet2010} who had demonstrated that particular classes of viscous flows between
porous rotating cylinders can be unstable to small two-dimensional perturbations.
Later it had been shown in \cite{IM2013a} that both converging and diverging inviscid irrotational flows can be linearly
unstable to two-dimensional perturbations and that the instability persists for high, but finite radial Reynolds numbers.
In \cite{IM2013b}, it had been shown that not only the particular examples of viscous steady flows considered in
\citep[][]{Gallet2010, IM2013a} can be unstable to two-dimensional perturbations,
but this is also true for much wider classes of both converging and diverging flows.
Further development of the two-dimensional theory can be found in a recent paper by \citet{Kerswell} where, among other things,
the effects of compressibility, three-dimensionality and nonlinearity have been considered. Kerswell has also pointed to a similarity
between the instability induced by the radial flow and the so-called stratorotational instability (SRI) which
is due to the axial density stratification in the Couette-Taylor flow (see also \cite{Gellert,Leclercq}).
The effect of three-dimensionality has been also studied in
\cite{IM2017}, where it has been shown that the basic flow is unstable provided that the radial Reynolds number is sufficiently
high and that, in almost all cases, the most unstable mode is two-dimensional, but never axisymmetric.
Finally, in a recent paper, \citet{Martinand2017} have studied the stability of the Couette-Taylor flow between porous cylinders
with both radial and axial flows. Among other things, they have found that the critical mode becomes non-axisymmetric when
the radial flow is sufficiently strong and the gap between the cylinders is sufficiently wide.

Even though the above papers resulted in a considerable progress in our understanding of the effect of a radial flow on the
stability of the Couette-Taylor flow, the full picture is still not very clear,
because of various restrictions that have been imposed either on the perturbations or on the basic flow.
The aim of the present paper is to fill most
of the existing gaps in the linear stability analysis and to obtain theoretical linear stability diagram similar to the diagram of
\citet{Andereck} that summarises the experimental results on the classical Couette-Taylor flow.

{We shall pay particular attention to the inviscid instability studied in \cite{IM2013a, IM2013b,Kerswell,IM2017}.
It was argued in \cite{IM2017} that the instability occurs because adding a radial flow to a circular flow between cylinders
radically changes the inviscid stability problem. One may think of the radial flow as a singular perturbation of the flow with circular streamlines. It is therefore not completely unexpected that adding the radial flow results in the appearance of new unstable
inviscid modes which do not exist in the absence of the radial flow. This is somewhat similar to
the tearing instability in the magnetohydrodynamics, where the
unstable tearing mode appears when a small resistivity is added
(see, e.g., \cite{Furth}). It had been shown in \cite{IM2017} that the inviscid instability persists for viscous flows,
provided that the radial Reynolds number is sufficiently high. For possible applications,
it is necessary to know whether this instability can survive for relatively low $R$, and
an answer to this question, which itself is important,  is the essential step towards our main aim.
}

The outline of the paper is as follows. {In Section 2, we state the governing equations, the basic flow and
the linearised stability problem. Section 3 describes the results of
the linear stability analysis. We start with a brief description of the numerical method. Then, in Sections 3.1 and 3.2
we consider the diverging flow in the case of a fixed outer cylinder ($\Rey_2=0$): Section 3.1 describes the behaviour of neutral
curves on the $(k,\Rey_1)$ plane ($k$ is the axial wave number of the perturbation) and Section 3.2 presents critical curves
on the $(\Rey_1,R)$ plane. In Section 3.3, we discuss the effect of rotation of the outer cylinder on the behaviour of
critical curves on the $(\Rey_1,R)$ plane. In Section 3.4, the converging flow with the inner cylinder fixed is considered and
critical curves on the $(\Rey_2,R)$ plane are discussed. {In Sections 3.5 and 3.6,
critical curves on the $(\Rey_2,\Rey_1)$ plane are presented: Section 3.6 deals with diverging flows and weak converging flows
and Section 3.5 covers strong converging flows}. Finally, discussion of the results is presented in Section 4.}


\section{Formulation of the problem}\label{sec:problem}

\subsection{Governing equations}

We consider three-dimensional viscous incompressible flows in the gap between two concentric circular cylinders
with radii $r_{1}$ and $r_{2}$ ($r_2 > r_1$). The cylinders are porous, and there is a constant volume flux $2\pi Q$
(per unit length along the common axis of the cylinders) of
the fluid through the gap (the fluid is pumped into the gap at the inner cylinder and taken out at the outer one or {\em vice versa}).
$Q$ is positive if the direction of the flow is from the inner cylinder to the outer one and negative if the flow direction is reversed.
Flows with positive and negative $Q$ will be referred to as diverging and converging flows respectively.
It is convenient to non-dimensionalise the Navier-Stokes equations using $r_1$ as a length scale, $r^2_{1}/\vert Q\vert$ as a time scale,
$\vert Q\vert/r_{1}$ as a scale for the velocity and $\rho Q^2/r_{1}^2$ as a scale
for the pressure. The dimensionless Navier-Stokes equations have the form
\begin{eqnarray}
&&u_{t}+ u u_{r} + \frac{v}{r}u_{\theta} + w u_{z} -\frac{v^2}{r}= -p_{r} +
\frac{1}{\vert R\vert} \left(\nabla^2 u-\frac{u}{r^2}-\frac{2}{r^2}v_{\theta}\right),  \label{N1} \\
&&v_{t}+ u v_{r} + \frac{v}{r}v_{\theta} + w u_{z} +\frac{u v}{r}= -\frac{1}{r} \, p_{\theta}
+ \frac{1}{\vert R\vert} \left(\nabla^2 v-\frac{v}{r^2}+\frac{2}{r^2}u_{\theta}\right),  \label{N2} \\
&&w_{t}+ u w_{r} + \frac{v}{r}w_{\theta} + w w_{z}  = - p_{z}
+ \frac{1}{\vert R\vert} \, \nabla^2 w ,  \label{N3} \\
&&\frac{1}{r}\left(r u\right)_{r} +\frac{1}{r} \, v_{\theta} + w_{z}=0.  \label{N4}
\end{eqnarray}
Here $(r,\theta,z)$ are the polar cylindrical coordinates, $u$, $v$ and $w$ are the radial, azimuthal and axial components
of the velocity, $p$ is the pressure, {$R = Q\rho/\mu$ is the radial Reynolds number  where $\mu$ is the dynamic viscosity of the fluid and $\rho$ is the  (constant) density};
subscripts denote partial derivatives; $\nabla^2$ is the polar form of the Laplace operator:
\[
\nabla^2=\pr_r^2 + \frac{1}{r}\pr_r + \frac{1}{r^2}\pr_\theta^2 + \pr_z^2.
\]
Note that the radial Reynolds number $R$ is zero without radial flow, positive for diverging flows and negative for converging flows.

We impose the following boundary conditions
\begin{equation}
u\!\bigm\vert_{r=1}=\beta, \quad u\!\bigm\vert_{r=a}=\frac{\beta}{a},
\quad v\!\bigm\vert_{r=1}=\gamma_1, \quad v\!\bigm\vert_{r=a}=\frac{\gamma_2}{a}, \quad w\!\bigm\vert_{r=1}=w\!\bigm\vert_{r=a}=0  \label{N5}
\end{equation}
where
\[
a=\frac{r_2}{r_1}, \quad \gamma_1=\frac{\Omega_1 r_1^2}{\vert Q\vert}, \quad
\gamma_2=\frac{\Omega_2 r_2^2}{\vert Q\vert}, \quad
\beta=\mathrm{sgn}(R)=\left\{
\begin{array}{ll}
R/\vert R\vert, &R\neq 0 \\
0,                      &R=0
\end{array}\right. .
\]
with $\Omega_1$ and $\Omega_2$ being the angular velocities of the inner and outer cylinders respectively.
Parameter $\beta$ takes values $+1$ and $-1$ for the diverging
and converging flows, respectively, and $0$ when there is no radial flow;
$\gamma_1$ and $\gamma_2$ represent the ratio of the azimuthal component of the velocity to the
radial one at the inner and outer cylinders  respectively.
Boundary condition (\ref{N5}) prescribe all components of the velocity at the cylinders and
model conditions on the interface between a fluid and a porous wall \citep[see, e.g.,][]{Joseph}.
In order to compare our results with existing literature on the Taylor-Couette flow, we shall
also use two azimuthal Reynolds numbers, $\Rey_1$ and $\Rey_2$, defined as
{
\[
\Rey_1=\frac{\rho\Omega_1 r_1(r_2-r_1)}{\mu}, \quad
\quad \Rey_2=\frac{\rho\Omega_2 r_2(r_2-r_1)}{\mu}.
\]}
Note that parameters $\gamma_1$ and $\gamma_2$ can be expressed in terms of $\Rey_1$ and $\Rey_2$ as
\[
\gamma_1 = \frac{1}{a-1} \, \frac{\Rey_1}{\vert R\vert }, \quad \gamma_2 = \frac{a}{a-1} \, \frac{\Rey_2}{\vert R\vert }.
\]
Problem (\ref{N1})--(\ref{N5}) has a rotationally and translationally invariant steady solution, given by
\begin{equation}
u=\frac{\beta}{r}, \quad v= V(r)=A r^{R +1} + \frac{B}{r}, \quad P=-\frac{1}{2r^2}+\int\frac{V^2(r)}{r} \, dr   \label{N6}
\end{equation}
where
\begin{equation}
A=\frac{\gamma_2 -\gamma_1}{a^{R+ 2}-1}, \quad B=\frac{a^{R + 2}\gamma_1 -\gamma_2}{a^{R + 2}-1}. \label{N7}
\end{equation}
{Steady solution (\ref{N6}), (\ref{N7}) is not defined for $R =-2$. In this special case, the solution is given by}
\begin{equation}
u=-\frac{1}{r}, \quad v= V(r)=\widetilde{A} \, \frac{\ln r}{r} + \frac{\widetilde{B}}{r}  \label{N8}
\end{equation}
where
\[
\widetilde{A}=\frac{\gamma_2 -\gamma_1}{\ln a}, \quad \widetilde{B}=\gamma_1.
\]
{\emph{Remark.} The non-dimensionalisation adopted above is very convenient for flows with a nonzero radial flux, but does not work for the classical
Couette-Taylor flow ($R=0$). To treat the latter, we simply re-scale the dimensionless quantities in (\ref{N1})--(\ref{N8}):
\[
t \rightarrow t/\gamma_1, \quad \mathbf{v} \rightarrow \gamma_1 \mathbf{v},  \quad p \rightarrow \gamma_1^2 p.
\]
This is equivalent to a non-dimensionalisation with $1/\Omega_1$ as the time scale, $r_1$ as the length scale, and $\Omega_1 r_1$ and
$\rho\Omega_1^2 r_1^2$ as the characteristic scales for the velocity and the pressure. With this re-scaling, the radial Reynolds number $\vert R\vert$ in (\ref{N1}) is replaced by $\Rey_1/(a-1)$. In the boundary conditions (\ref{N5}), $\beta \rightarrow \beta/\gamma_1$, $\gamma_1 \rightarrow 1$ and
$\gamma_2\rightarrow \gamma_2/\gamma_1$. If $\gamma_1=0$, a similar re-scaling can be done with $\gamma_1$ replaced by $\gamma_2$, but this case is not relevant
for the present paper.}

{Returning to the discussion of the basic flow (\ref{N6})--(\ref{N8}), we note that the azimuthal velocity profile has a non-trivial dependence on $R$. When $R=0$ (no radial flow), the velocity
(re-scaled in accordance with the above Remark) reduces to the classical Couette-Taylor profile,
$V(r)/\gamma_1= (Ar+B/r)/\gamma_1$.} When $R\to\infty$, the limit of $V(r)$ depends on the sign of $R$, i.e. on the direction of the radial flow.
It can be shown \citep[see][]{IM2013b} that,
for $\vert R\vert\gg 1$, the azimuthal component of the velocity is well approximated by
\begin{equation}
V(r)=
\left\{
\begin{array}{ll}
\gamma_1/r + f(\eta)/a + O\left(R^{-1}\right) &\hbox{for} \ \ \beta=1 \\
\gamma_2/r - f(\xi) + O\left(R^{-1}\right) &\hbox{for} \ \ \beta=-1
\end{array}\right. \label{N9}
\end{equation}
where $\eta=\vert R\vert (1-r/a)$ and $\xi=\vert R\vert (r-1)$ are the boundary layer variables (at the outer and inner cylinders respectively)
and function $f$ is defined as
\begin{equation}
f(s)=(\gamma_2 - \gamma_1) \, e^{-s}. \label{N10}
\end{equation}
Equations (\ref{N9}) and (\ref{N10}) mean that, in the limit of high Reynolds numbers,
the flow becomes \emph{irrotational and proportional to $r^{-1}$ everywhere except for
a thin boundary layer near the outflow part of the boundary} (i.e. near the outer cylinder for the diverging flow and the inner
cylinder for the converging flow). The boundary layer thickness is $O(\vert R\vert^{-1})$. Note that there is no boundary layer at
the inflow part of the boundary. This is consistent with
the general theory of flows through a given domain with permeable boundary in the limit of vanishing viscosity
(see, e.g., \cite{Temam,Yudovich2001,Ilin2008,Korobkov}).

If the boundary layer is ignored, we obtain the corresponding inviscid flow:
\begin{equation}
u=\frac{\beta}{r}, \quad v=
\left\{
\begin{array}{ll}
\gamma_1/r &\hbox{for} \ \ \beta=1 \\
\gamma_2/r &\hbox{for} \ \ \beta=-1
\end{array}\right. \label{N11}
\end{equation}
Note that the single inviscid flow (\ref{N11}) represents the high-Reynolds-number limit for each member
of a one-parameter family of viscous flows (\ref{N6}) (parametrised by $\gamma_2$ for $\beta=1$ and by $\gamma_1$ for $\beta=-1$).

The inviscid flow (\ref{N11}) is the only steady rotationally symmetric and translationally invariant (in the $z$
direction) solution of the Euler equations
that satisfies the boundary conditions
\begin{equation}
u\!\bigm\vert_{r=1}=1, \quad u\!\bigm\vert_{r=a}=\frac{1}{a},
\quad v\!\bigm\vert_{r=1}=
\gamma_1, \quad w\!\bigm\vert_{r=1}=0  \label{N12}
\end{equation}
for the diverging flow and
\begin{equation}
u\!\bigm\vert_{r=1}=-1, \quad u\!\bigm\vert_{r=a}=-\frac{1}{a},
\quad v\!\bigm\vert_{r=a}=
\frac{\gamma_2}{a}, \quad w\!\bigm\vert_{r=a}=0  \label{N13}
\end{equation}
for the converging flow. In both cases, the boundary conditions at the inflow part
of the boundary include all components of the velocity (not only the normal one as in the case of impermeable boundary).
These boundary conditions are special ones because (i) they lead to a well-posed mathematical problem
(see, e.g., \cite{Monakh}) and (ii) they are consistent with the vanishing viscosity limit for the Navier-Stokes equations
(see, e.g., \cite{Temam, Ilin2008}). It should also be mentioned that other types of
boundary condition can be employed for inviscid flows through a domain with permeable boundary
(e.g., \cite{Monakh, MorgYud}). Some of these alternative
boundary conditions lead to mathematically well-posed problems. However, only conditions (\ref{N12}) and (\ref{N13})
are consistent with the vanishing viscosity limit for the Navier-Stokes equations (\ref{N1})--(\ref{N4}) with boundary conditions
(\ref{N5}).

Before we turn our attention to the stability analysis, it is useful to recall some stability properties of the classical Couette-Taylor flow
and to discuss their relevance for the steady flow (\ref{N6}).  The classical Couette-Taylor flow ($R=0$) is
centrifugally unstable to inviscid axisymmetric perturbations if the Rayleigh discriminant, given by
$\Phi(r)=r^{-3}d(rV(r))^2/dr$,
is negative somewhere in the flow and stable if $\Phi(r) > 0$ for all $1< r <a$ (see, e.g., \cite{Chandra, Drazin}).
It should be noted that although there is no evidence
suggesting that the Couette-Taylor flow may be unstable to non-aixisymmetric perturbations if $\Phi(r) > 0$ for all $1< r <a$,
the stability has never been formally proved, except for the case of large axial wavenumbers \cite{Billant}.
According to the Rayleigh criterion, the classical Couette-Taylor flow is always unstable if
$\gamma_1$ and $\gamma_2$ have different signs. For positive $\gamma_1$ and $\gamma_2$, the regions of stable and unstable Couette-Taylor flows are separated by the Rayleigh line, $\gamma_2=\gamma_1$ (or, equivalently, $\Rey_1= a\, \Rey_2$).
Although the Couette-Taylor flow with a radial flow, given by (\ref{N6}), is different from the classical
Couette-Taylor flow,
the Rayleigh discriminant has the same properties for any $R\neq 0$: $\Phi(r)>0$ (for $0<r<a$) if $\gamma_1<\gamma_2$,
$\Phi(r)<0$ (for $0<r<a$) if $\gamma_1>\gamma_2$, and $\Phi(r)\equiv 0$ for $\gamma_1=\gamma_2$. However, as has been shown in \cite{IM2017},
the presence of the radial flow radically changes the stability properties of the Couette-Taylor flow: any flow with sufficiently
large $\gamma_1$ and $\gamma_2$ turns out to be unstable in the limit of high radial Reynolds numbers irrespective of whether
$\gamma_1$ is smaller or larger than $\gamma_2$. This means that the Rayleigh
criterion is not relevant for the basic flow (\ref{N6}) at least when $\vert R\vert\gg 1$.

\subsection{Linear stability problem}

Let a small perturbation
$(\tilde{u}, \tilde{v}, \tilde{w}, \tilde{p})$ of the basic flow (\ref{N6}) have the form of the normal mode
\begin{equation}
\{\tilde{u}, \tilde{v}, \tilde{w}, \tilde{p}\} = Re\left[\{\hat{u}(r), \hat{v}(r), \hat{w}(r), \hat{p}(r)\} e^{\sigma t + in\theta+ ikz}\right]  \label{N14}
\end{equation}
where $n\in\mathbb{Z}$ and $\sigma\in\mathbb{C}$. This leads to the eigenvalue problem for $\sigma$:
\begin{eqnarray}
&&\left(\sigma +  \frac{in V}{r} + \frac{\beta}{r} \, \pr_r \right) \hat{u}
-\frac{\beta}{r^2} \, \hat{u} -\frac{2V}{r} \, \hat{v} = - \pr_r \, \hat{p}  +
\frac{1}{\vert R\vert} \left(L \hat{u}-\frac{\hat{u}}{r^2}-\frac{2in}{r^2} \, \hat{v}\right) ,  \label{N15} \\
&&\left(\sigma +  \frac{in V}{r} + \frac{\beta}{r} \, \pr_r \right) \hat{v}
+\frac{\beta}{r^2} \, \hat{v}  +\Omega(r) u = -\frac{in}{r} \, \hat{p}  +
\frac{1}{\vert R\vert} \left(L \hat{v} -\frac{\hat{v}}{r^2}+\frac{2in}{r^2} \, \hat{u}\right),  \label{N16} \\
&&\left(\sigma +  \frac{in V}{r} + \frac{\beta}{r} \, \pr_r \right) \hat{w}
 = -ik \, \hat{p}  +
\frac{1}{\vert R\vert} \, L \hat{w} ,  \label{N17} \\
&&\pr_r \left(r \hat{u}\right) +in \, \hat{v} + ikr \, \hat{w}=0, \label{N18} \\
&&\hat{u}(1)=0, \quad \hat{u}(a)=0, \quad \hat{v}(1)=0, \quad \hat{v}(a)=0, \quad \hat{w}(1)=0, \quad \hat{w}(a)=0.  \label{N19}
\end{eqnarray}
In Eqs. (\ref{N15})--(\ref{N19}),
\[
L  = \frac{d^2}{dr^2} + \frac{1}{r} \, \frac{d}{dr} - \left(k^2 + \frac{n^2}{r^2}\right), \quad \Omega(r)=V'(r)+\frac{V(r)}{r} .
\]
If there is an eigenvalue $\sigma$ such that
$Re(\sigma)>0$, then the basic flow is unstable. If there are no eigenvalues with positive real part and if there are no
perturbations with non-exponential growth, then it is linearly stable. Although the possibility of  the non-exponentially
growing perturbations certainly deserves attention especially in the vanishing viscosity limit, this question is beyond the scope
of this paper. For $k=0$, problem (\ref{N15})--(\ref{N19}) reduces to the two-dimensional viscous stability problem that had
been studied in \cite{IM2013b}.

Eigenvalues $\sigma$ depend on the azimuthal and axial wavenumbers, $n$ and $k$, i.e. $\sigma=\sigma(n,k)$. It is easy to see that
(i) if $\sigma(n,k)$ is an eigenvalue with eigenvector $(\hat{u}, \hat{v}, \hat{w}, \hat{p})$, then $\sigma(-n,-k)=\sigma^*(n,k)$
is also an eigenvalue with eigenvector $(\hat{u}^*, \hat{v}^*, \tilde{w}^*, \hat{p}^*)$ (where star denotes complex conjugation) and
(ii) if $\sigma(n,k)$ is an eigenvalue with eigenvector $(\hat{u}, \hat{v}, \hat{w}, \hat{p})$, then $\sigma(n,-k)=\sigma(n,k)$ is
also an eigenvalue with eigenvector
$(\hat{u}(n,-k), \hat{v}(n,-k), \hat{w}(n,-k), \hat{p}(n,-k))=(\hat{u}(n,k), \hat{v}(n,k), -\hat{w}(n,k), \hat{p}(n,k))$.
The latter property implies that all eigenvalues with $k\neq 0$ are double with two independent eigenvectors
\[
\left.(\hat{u}, \hat{v}, \tilde{w}, \hat{p})\right\vert_{n,k}e^{in\theta+ ikz}, \quad
\left.(\hat{u}, \hat{v}, -\hat{w}, ~\hat{p})\right\vert_{n,-k} e^{in\theta- ikz}.
\]
So, in this respect, the situation is exactly the same as that for the classical Couette-Taylor flow (see, e.g., \cite{Iooss}).


\section{Linear stability analysis}


Eigenvalue problem (\ref{N15})--(\ref{N19}) was solved numerically using an adapted version of
a Fourier-Chebyshev Petrov-Galerkin spectral method described by \citet{Trefethen2003}.
We have computed the eigenvalue with largest real part, $\sigma$, for a range of values of the radial and azimuthal Reynolds numbers $R$, $\Rey_1$ and $\Rey_2$
and for three
different values of the geometric parameter $a$ ($a=2$, $4$ and $8$).
The neutral curves ($\Rey_1(k)$ for fixed $R$ and $\Rey_2$, and $\Rey_2(k)$  for fixed $R$ and $\Rey_1$)
were computed using the secant method. Critical values of $\Rey_1$ (or $\Rey_2$), $n$ and $k$ were obtained by minimization of neutral curves over
all $n\geq 0$ and $k\geq 0$. [The minimization has been performed over a finite number of azimuthal modes and over a finite interval in $k$,
but we have no doubt that the results would remain unchanged, if we were able to do this for all $n\geq 0$ and $k\geq 0$.]
Minimization over $k$ was done using the golden section search algorithm (see, e.g., \cite{Press}).

The numerical method has been validated (i) by computing critical values of $\Rey_1$ in the case of $\Rey_2 = 0$ and relatively small values of
$\vert R\vert$ ($\vert R\vert\lesssim 15$) and comparing them with the known results of \cite{Min, Johnson} (see section 3.2 below) and (ii) by comparing our numerical curves
for $\vert R\vert \gg 1$ with relevant results of the inviscid theory \cite{IM2017} (see sections 3.2 and 3.3 below).

Out aim is to determine the stability/instability regions in the plane of parameters $\Rey_1$ and $\Rey_2$. The properties of the eigenvalues stated above imply that we only need to consider modes with non-negative $n$ and $k$.


{\subsection{Examples of neutral curves for diverging flows with non-rotating outer cylinder}}

\begin{figure}
\begin{center}
\includegraphics*[height=11cm]{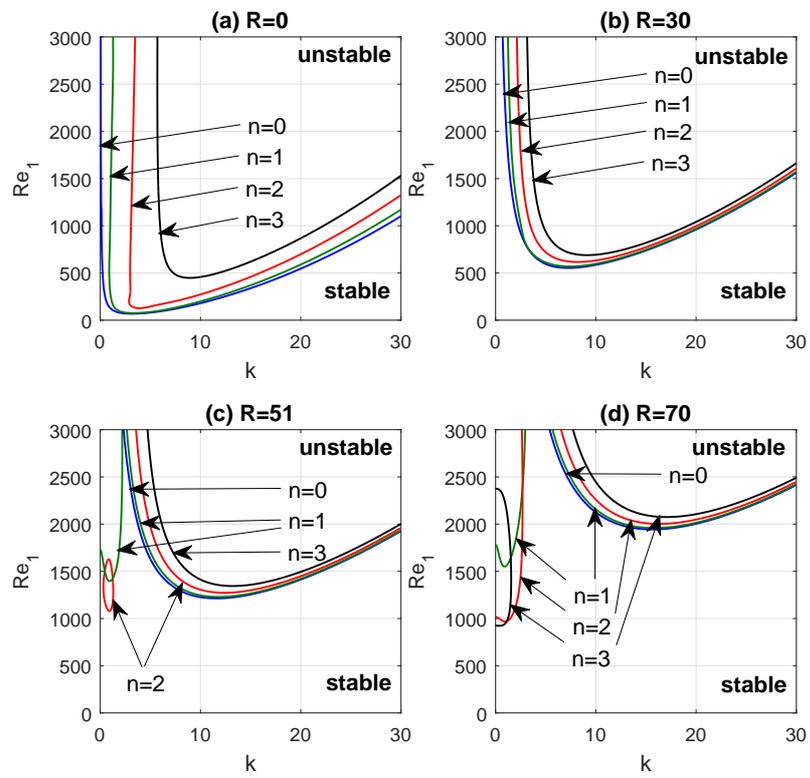}
\end{center}
\caption{Neutral curves for the first 4 azimuthal modes ($n=0,\dots,3$), $a=2$ and $\Rey_2=0$: (a) - $R=0$; (b) - $R=30$; (c) - $R=50$; (d) - $R=70$.}
\label{IM_fig1}
\end{figure}

To understand what happens when a radial flow is added to the classical Couette-Taylor flow, we first consider the non-rotating outer cylinder
(i.e. $\Rey_2=0$). Neutral curves ($\Real(\sigma) = 0$) for several values of the radial Reynolds number are shown in Fig. \ref{IM_fig1}(a,b,c,d).
The neutral curves for the classical Couette-Taylor flow ($R=0$) are shown in Fig. \ref{IM_fig1}(a). When the azimuthal Reynolds number $\Rey_1$ increases
from zero, the mode that becomes unstable first is an axisymmetric mode ($n=0$) for any value of the axial wave number $k\geq 0$. By minimizing $\Rey_1$ over all $n$ and $k$, we can obtain the critical Reynolds number $\Rey_{1cr}$ and the corresponding values of $n$ and $k$: $\Rey_{1cr}\approx 68.2$, $n_{cr}=0$ and $k_{cr}\approx 3.16$ in Fig. \ref{IM_fig1}(a). Figure \ref{IM_fig1}(b) displays the neutral curves for $R=30$. One can see that
the curves keep qualitatively the same shape (as those in \ref{IM_fig1}(a)), but are shifted upwards (to higher values of $\Rey_{1}$) and to the right (to the region of
shorter axial waves). The critical values of $\Rey_{1}$, $n$ and $k$ in Fig. \ref{IM_fig1}(b) are: $\Rey_{1cr}\approx 553.5$, $n_{cr}=0$ and $k_{cr}\approx 7.24$. As
$R$, the Couette-Taylor modes shift further up and to the right. This however is not the end of the story. When the radial Reynolds number becomes  sufficiently high, another neutral curve for the mode $n=1$ (disconnected from the corresponding Couette-Taylor mode)
appears in the long wave region of the $(k,\Rey_1)$ plane. {This occurs at $R\approx 39.6834$. The new curve is closed and
encircles a small oval instability region with a center approximately at $(k,\Rey_1)=(0.84, 1974)$.
As $R$ increases further, the instability region for mode $n=1$ grows and becomes attached to
the vertical axis ($k=0$). Then (at $R\approx 49.953$), another oval instability region for the mode $n=2$ appears. At first the instability regions for modes $n=2$ and $n=1$ do not intersect, but as $R$ increases, they grow in size and overlap.}
Figure \ref{IM_fig1}(c) shows the neutral curves for $R=51$. One can see that, in addition to the unstable region
bounded by the axisymmetric Couette-Taylor mode, there are two intersecting unstable regions for modes $n=1$ and $n=2$ corresponding to long axial waves.
The first of these is pretty large and attached to the vertical axis ($k=0$), while the second is a relatively small oval region.
This oval region will grow with $R$ and also become attached to the vertical axis at higher values of $R$. The most unstable mode (the critical mode)
in Fig. \ref{IM_fig1}(c)
(obtained by minimizing $\Rey_1$) is the non-axisymmetic mode with $\Rey_{1cr}\approx 1077$, $n_{cr}=2$ and $k_{cr}\approx 0.91$.
Neutral curves for $R=70$ are shown in Fig. \ref{IM_fig1}(d). Now there are three non-axisymmetric modes ($n=1,2,3$) that are unstable to perturbations with sufficiently
small $k$. The most unstable mode is $n_{cr}=3$, and the corresponding critical Reynolds number and the axial wave number are $\Rey_{1cr}\approx 923.2$
and $k_{cr}\approx 0.384$. {For $R=70$, all modes with $n>3$ remain stable.}

Two conclusions can be drawn from the above observations: (i) the diverging radial flow forces the classical Couette-Taylor modes to shift from the region
of relatively low
$\Rey_1$ to the region of higher
$\Rey_1$ and higher axial wave numbers; (ii) new spiral long-wave unstable modes appear
when the radial Reynolds number increases and those become more unstable than the axisymmetric Couette-Taylor mode. These non-axisymmetric modes are what is left of inviscid unstable modes (see
\cite{IM2013a, IM2017, Kerswell}) after the viscous effects were switched on.


{\subsection{Critical curves for converging and diverging flows with non-rotating outer cylinder}}

\begin{figure}
\begin{center}
\includegraphics*[height=10cm]{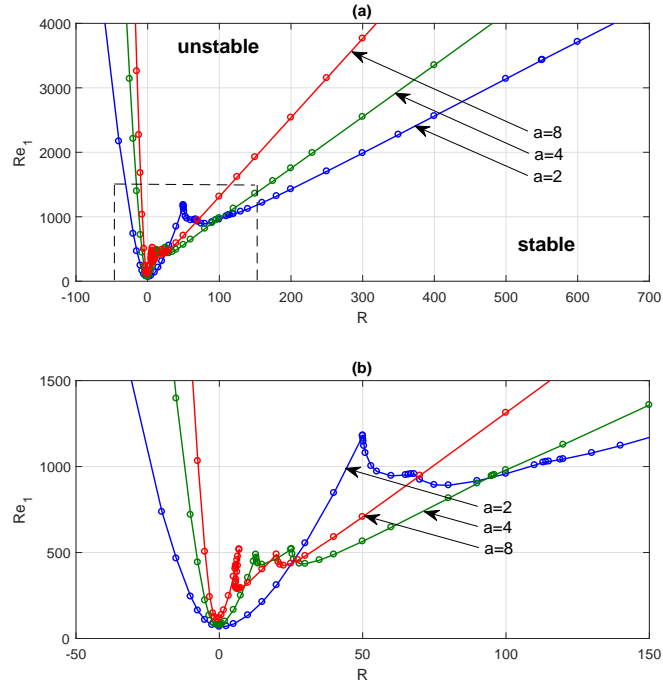}
\end{center}
\caption{Critical curves for the diverging flow with the fixed outer cylinder ($\Rey_2=0$) for $a=2, 4, \ \hbox{and} \ 8$. Circles show the computed points; solid curves are obtained by linear spline interpolation. Figure (b) is a magnified rectangular region (bounded by dashed lines) of figure (a).}
\label{IM_fig2}
\end{figure}

By computing neutral curves and then the critical azimuthal Reynolds number, $\Rey_{1cr}$ and the corresponding wave numbers $n_{cr}$ and $k_{cr}$
for different values of the radial Reynolds number, we can obtain stability/instability regions in the $(R, \Rey_1)$ plane,
separated by critical curves: $\Rey_1 = \Rey_{1cr}(R)$.
These are shown in Fig. \ref{IM_fig2} for $a=2,4 \ \hbox{and} \ 8$.
\begin{figure}
\begin{center}
\includegraphics*[height=9cm]{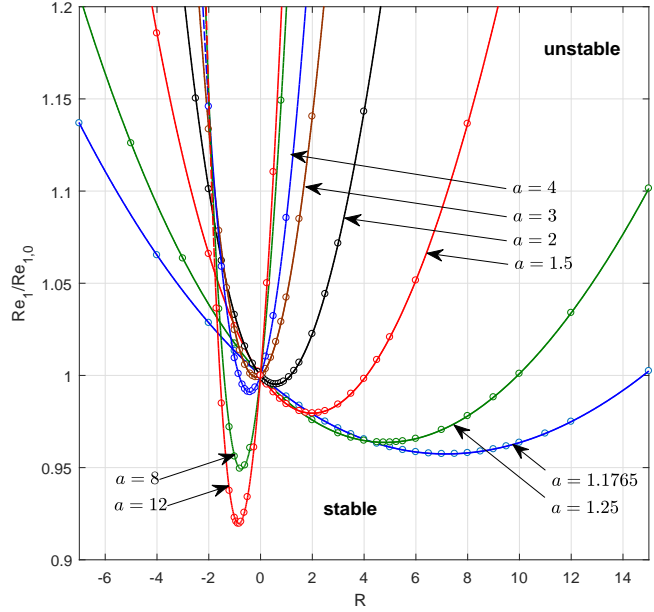}
\end{center}
\caption{Critical values of $\Rey_1$ (normalized by $\Rey_{1,0}$, the critical value of $\Rey_1$ at $R=0$) for $-7< R < 15$ and various values of $a$.}
\label{NEAR_ZERO}
\end{figure}
In Fig. \ref{IM_fig2}, circles represent the computed points, and solid curves are obtained by linear spline interpolation. Figure \ref{IM_fig2}(b) is obtained by
enlarging the rectangular part of Fig. \ref{IM_fig2}(a) (bounded by dashed lines). Nevertheless,
there is a subtle effect which is not seen even in Figure \ref{IM_fig2}(b): the global minimum of the critical curve
for $a=2$ is attained not at $R=0$ but
at $R\approx 0.56$ ($\Rey_1\approx 67.86$), so that a very weak diverging radial flow has a small destabilising effect. This agrees with
what had been found earlier by \citet{Min} for $a$ in the range $[1.05, 2]$ (see also
\cite{Johnson}).

{To better understand the dependence of the stability properties of the flow on the gap size,
we have considered relatively weak radial flows ($|R|<15$)
in more detail}, paying particular attention to wider gaps ($a>2$).
Critical curves for several values of $a$, normalized by $\Rey_{1,0}$, the critical value of $\Rey_1$ at $R=0$ (no radial flow), are shown in Fig. \ref{NEAR_ZERO}.
Curves for $a \leq 2$ agree with the corresponding curves in \cite{Min,Johnson}
(and this gives additional validation to our numerical computations).
Figure \ref{NEAR_ZERO} shows that the global minimum shifts to the left, when $a$ increases, and the minimum is
attained at $R=0$ for $a\approx 2.64$ and at negative $R$ for wider gaps, $a > 2.64$.
This means that the effect of a weak radial flow for $a > 2.64$
is opposite to that for $a < 2.64$. In other words,
a weak converging radial flow has a destabilizing effect, while a weak diverging flow is stabilising for $a> 2.64$.
This fact appears to have been missed in previous studies.

\begin{figure}
\begin{center}
\includegraphics*[height=9cm]{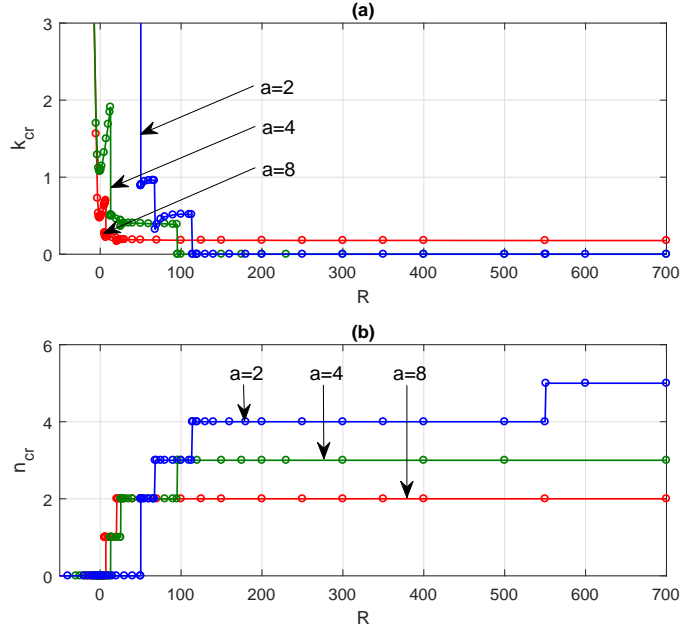}
\end{center}
\caption{Critical values of azimuthal and axial wave numbers, $n_{cr}$ and $k_{cr}$. Circles show the computed points; solid curves are obtained by linear spline interpolation. Figure (a) shows $k_{cr}$ versus $R$ and figure (b) - $n_{cr}$ versus $R$.}
\label{IM_fig3}
\end{figure}

Now we return to the discussion of the global picture. Figure \ref{IM_fig2} shows that when the radial Reynolds number
decreases from its value at the global minimum, the critical values of $\Rey_1$
monotonically increase for all values of $a$ considered in the present paper, and for sufficiently strong converging radial flows ($R < 0$),
the growth of the critical Reynolds number $\Rey_1$ becomes quite rapid as $R$ decreases.
The critical mode remains axisymmetric and the critical axial wave number increases (the critical wave numbers,
$n_{cr}$ and $k_{cr}$, as functions of $R$ are shown in Fig. \ref{IM_fig3}(a,b)).
When the radial Reynolds number increases from its value at the global minimum, the behaviour of the critical curves is more interesting.
At first, over a finite interval (in $R$), the critical Reynolds number, $\Rey_{1cr}$, increases. Then, at a certain value of $R$,
it starts decreasing, attains a local minimum and then increases over another finite interval up to another value of $R$, at which it starts decreasing again.
Such behaviour may be repeated several times depending on the thickness of the gap.
Each transition from growth to decay occurs at a point where the smoothness of the
critical curve is lost (see Fig. \ref{IM_fig2}) due to switching to a critical mode with different azimuthal and axial wave numbers.

{Critical values of the axial and azimuthal wave numbers, $k_{cr}$ and $n_{cr}$, as functions of $R$ are
shown in Fig. \ref{IM_fig3}.
The jumps in $k_{cr}$ occur when the azimuthal wave number, $n_{cr}$, of the most unstable mode changes. The number
of transitions from one value of $n_{cr}$ to another depends on the gap between the cylinders: it is smaller for larger gaps.
In all transitions in Fig. \ref{IM_fig3}(b), $n_{cr}$ increases by $1$, except for the first transition for $a=2$ where the azimuthal wave number jumps from $n_{cr}=0$ to $n_{cr}=2$.}

The above observations are
in agreement with the results of \citet{Martinand2017} where
the appearance of non-axisymmetric unstable modes have been found for $a=4$ (see Fig. 4 in \cite{Martinand2017}).
{For large $R$ ($R>600$ in Fig. \ref{IM_fig3}), the critical azimuthal wave number remains unchanged, while
the critical value of the axial wave number approaches a limit. Both the critical azimuthal wave number and the limit of
the critical axial wave number coincide with values provided by
the inviscid theory \cite{IM2017} (and this gives one more validation for our numerical results).}
For sufficiently large $R$, the most unstable modes are two-dimensional modes
(i.e. $k_{cr}\to 0$ as $R\to\infty$) with $n_{cr}=5$ for $a=2$ and $n_{cr}=3$ for $a=4$,
and this observation is consistent with the inviscid theory (see \cite{IM2017}). For
$a=8$, the azimuthal wave number of the most unstable mode is $n_{cr}=2$, while the critical axial wave number does not vanish in the limit $R\to\infty$, rather,
it tends to a finite limit: $k_{cr}\to 0.177$ as $R\to\infty$. This is also in agreement with the inviscid theory.
Note that this fact has been missed
in \cite{IM2017} where it was reported that the most unstable inviscid mode is two-dimensional for all $a$ ($a=2, 4 \ \hbox{and} \ 8$).

Another connection of the stability/instability regions in Fig. \ref{IM_fig2} with the inviscid theory is that, for large $R$, the slope of each curve
tends to a limit determined by the inviscid theory, namely:
\[
\frac{\Rey_{1cr}}{R} = (a-1) \, \gamma_1 \ \to \ (a-1) \, \gamma_{div}(a) \quad \hbox{as} \quad R\to\infty
\]
where $\gamma_{div}(a)$ is the critical value of $\gamma_1$ for the diverging flow in the inviscid theory (see \cite{IM2017}): $\gamma_{div}(2)\approx 5.73$,
$\gamma_{div}(4)\approx 2.68$ and $\gamma_{div}(8)\approx 1.75$. Two more interesting features can be seen in Figs. \ref{IM_fig2} and \ref{IM_fig3}:
(i) the value of the radial Reynolds number, at which the first transition from
an axisymmetric critical mode to a non-axisymmetric one occurs, is smaller for larger values of $a$ and (ii) for sufficiently large $R$,
the azimuthal wave number of the most unstable mode is higher for smaller $a$. We note in passing that for larger values of $a$ (beyond $a=8$),
the azimuthal wave number of the most unstable mode seems to remain unchanged: $n_{cr}=2$.

The main conclusions that can be drawn from Figs. \ref{IM_fig2}, \ref{NEAR_ZERO} and \ref{IM_fig3} are: (i) for $a > 2.64$, i.e. for a large
gap between the cylinders, a weak diverging flow has a stabilising effect on the Couette-Taylor flow, while a weak converging flow is destabilising, which is the opposite effect to what happens for relatively narrow gaps; (ii) the non-axisymmetric unstable modes, originated from the inviscid instability
(see \cite{IM2013a, IM2017, Kerswell}), can be observed at moderate values of the radial Reynolds number:
$R\gtrsim 50$ for $a=2$, $R\gtrsim 12.8$ for $a=4$ and $R\gtrsim 7$ for $a=8$.


{\subsection{Critical curves for converging and diverging flows when the outer cylinder is rotating}}

\begin{figure}
\begin{center}
\includegraphics*[height=9cm]{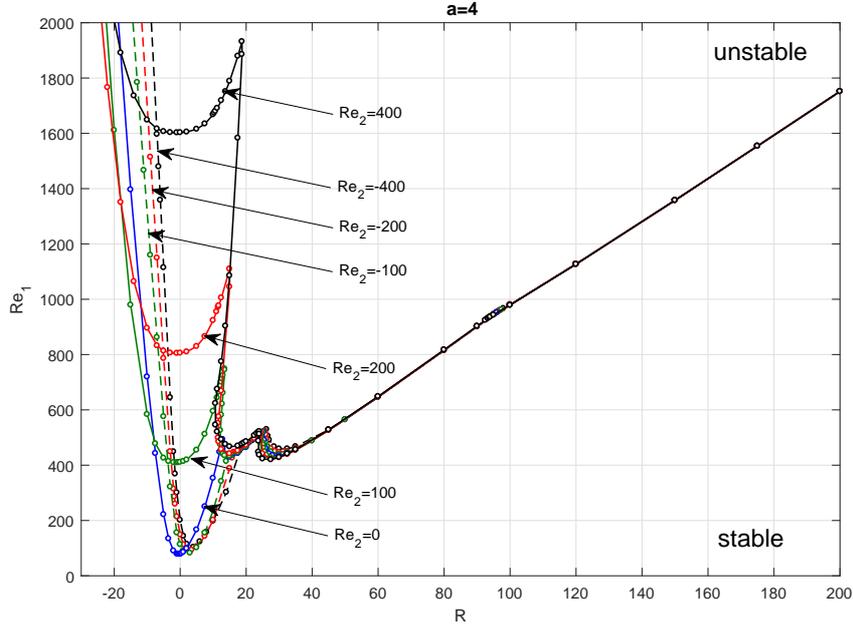}
\end{center}
\caption{Critical curves for the converging and diverging flows for $a=4$ and $\Rey_2=-400, -200, -100, 0, 100, 200, 400$. Circles show the computed points; solid curves are obtained by linear spline interpolation.}
\label{NEW_FIG1}
\end{figure}

{So far, we have looked at a particular class of the diverging and converging flows with fixed outer cylinder. Now we examine the effect of the rotation
of the outer cylinder on the stability of the basic flow. To this end, we have computed the critical curves
on the $(R,\Rey_1)$ plane for several nonzero values of $\Rey_2$ for $a=4$. These are shown on Fig. \ref{NEW_FIG1}.}
{Evidently, when the outer cylinder is rotating in the opposite direction relative to the inner one ($\Rey_2 < 0$), the critical curves slightly shift to the right, but their shape remains very similar to the curve for $\Rey_2=0$. When the cylinders are rotating in the same direction ($\Rey_2 > 0$), the part of the curve near the vertical line $R=0$ rapidly shifts up (with $\Rey_2$), because the classical Couette-Taylor flow (without the radial flow) becomes more stable. Interestingly,
the part of the critical curve corresponding to the inviscid instability changes very little even for moderate positive values of $R$.
For sufficiently large $R$ ($R>100$), the critical curves for $\Rey_2\in [-400,400]$ are almost indistinguishable. This means that the stability properties of the diverging flow are almost independent of whether or not
the outer cylinder is rotating, and this is consistent with earlier results reported in \cite{IM2013b}.}

\begin{figure}
\begin{center}
\includegraphics*[height=10cm]{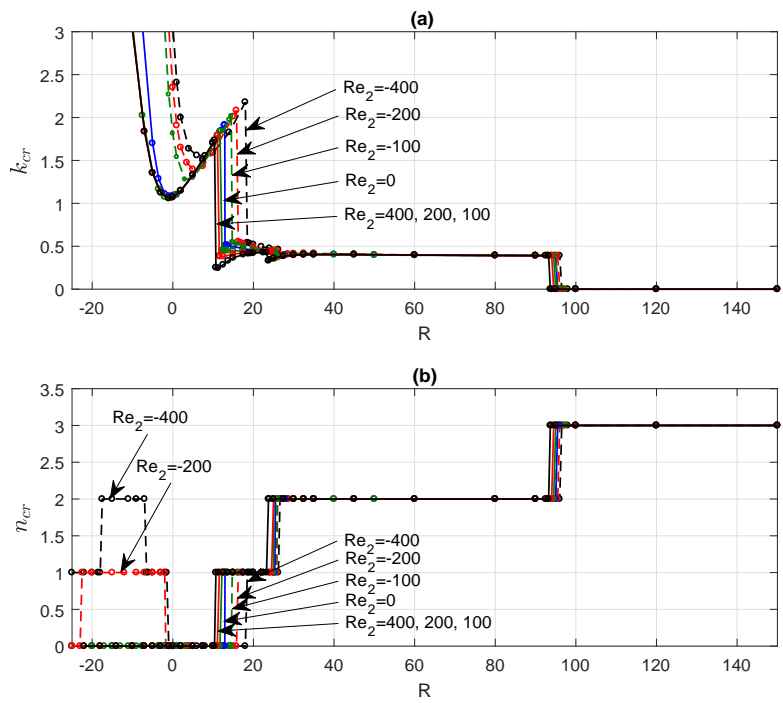}
\end{center}
\caption{Critical values of the axial and azimuthal wave numbers, $k_{cr}$ and $n_{cr}$, versus $R$ for $a=4$ and $\Rey_2=-400, -200, -100, 0, 100, 200, 400$. Circles show the computed points; solid curves are obtained by linear spline interpolation.}
\label{NEW_FIG2ab}
\end{figure}

{The critical values of the axial and azimuthal wave numbers as functions of $R$ for $a=4$ and several
fixed values of $\Rey_2$ are presented in Fig. \ref{NEW_FIG2ab}.
Figure \ref{NEW_FIG2ab}(a) shows $k_{cr}$ versus $R$; the jumps in $k_{cr}$ correspond to transitions between modes
with different azimuthal wave numbers (see Fig. \ref{NEW_FIG2ab}(b)). The qualitative behaviour of the curves, $k_{cr}(R)$, remains
very similar when $\Rey_2$ is varied in the interval $[-400,400]$;
the curves for positive $\Rey_2$ shift slightly to the left, while those for
negative $\Rey_2$ shift to the right. For all $\Rey_2\in[-400,400]$ and for sufficiently large positive
$R$ ($R\geq 100$), the critical value of the axial wave number remains equal to zero, with the critical azimuthal wave number being equal to $3$. Thus, for sufficiently large positive $R$, the most unstable mode is two-dimensional ($k_{cr}=0$)
with the azimuthal wave number $n_{cr}=3$, and this is in agreement with the inviscid theory \cite{IM2017}. }

{The behaviour of the critical azimuthal wave number, $n_{cr}$, for several values of $\Rey_2$ is shown in Fig. \ref{NEW_FIG2ab}(b).
For $\Rey_2=-100,100, 200$ and $400$, it is similar to that for $\Rey_2=0$. For $\Rey_2=-200$ and $-400$, $n_{cr}$ becomes nonzero for the converging flow ($R<0$). One can see in Fig. \ref{NEW_FIG2ab}(b) that $n_{cr}$ may be
equal to $1$ for $\Rey_2=-200$, and to $1$ and $2$ for $\Rey_2=-400$. This, however, is not very surprising because similar results are
well-known for the classical Couette-Taylor flow (see e.g. \cite{Iooss}): the most unstable mode becomes non-axisymmetric when the cylinders rotate in opposite directions with sufficiently high angular velocities, i.e. for $\Rey_2<0$ and $\vert\Rey_2\vert \gg 1$. For $a=4$, the transition from $n_{cr}=0$ to $n_{cr}=1$ in the classical Couette-Taylor flow occurs at $\Rey_2\approx -1050$. Figure \ref{NEW_FIG2ab}(b) shows the converging radial flow forces such transition at $\Rey_2$ as low as $200$.
 }


{\subsection{Critical curves for converging flows with non-rotating inner cylinder}}

\begin{figure}
\begin{center}
\includegraphics*[height=11cm]{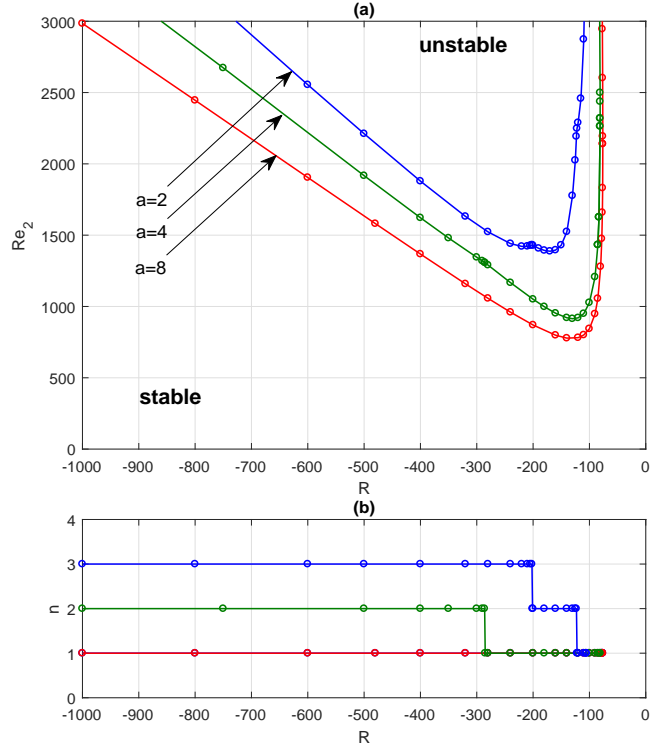}
\end{center}
\caption{Figure (a): Critical curves for the converging flow with the fixed inner cylinder ($\Rey_1=0$) for $a=2, 4, \ \hbox{and} \ 8$. Circles show the computed points; solid curves are obtained by linear spline interpolation. Figure (b):  critical azimuthal wave number, $n_{cr}$, versus radial Reynolds number, $R$.}
\label{IM_fig4}
\end{figure}

Now we consider the situation when the inner cylinder is fixed. It is well-known that the classical Couette-Taylor flow is stable in this case.
If we impose a diverging radial flow ($R>0$), the basic flow remains stable. However, a converging radial flow can be destabilising.
Figure \ref{IM_fig4}(a) shows the stability/instability regions in the $(R, \Rey_2)$ plane for $\Rey_1 =0$ and $a=2, 4 \ \hbox{and} \ 8$. Azimuthal wave number of the
most unstable mode is shown in Fig. \ref{IM_fig4}(b). As before, the circles represent the computed points; the solid curves are obtained by linear spline interpolation.
In all cases, the axial wave number of the most unstable mode is zero (i.e. the most unstable mode is a two-dimensional azimuthal wave).
The critical curves for $a=2$ and $a=4$ have points
where the derivative along the curve is discontinuous (this is
similar to what we had for the diverging flow in the preceding subsection). These points are where the azimuthal wave number of the most unstable mode changes.
There are no such points on the critical curve for $a=8$: it turns out that $n_{cr}=1$ for all relevant values of $R$.
For sufficiently strong converging flows ($R<0$ and $\vert R\vert \gg 1$),
both the values of the critical azimuthal wave number and the slopes of the curves agree with the inviscid results of \cite{IM2017}. In particular,
\[
\frac{\Rey_{2cr}}{R} = -\frac{a-1}{a} \, \gamma_2 \ \to \ -\frac{a-1}{a} \, \gamma_{conv}(a) \quad \hbox{as} \quad R\to -\infty
\]
where $\gamma_{conv}(a)$ is the critical value of $\gamma_2$ for the diverging flow in the inviscid theory (see \cite{IM2017}): $\gamma_{conv}(2)\approx 7.16$,
$\gamma_{conv}(4)\approx 4.21$ and $\gamma_{conv}(8)\approx 3.15$.

The above observations can be summarised as follows: (i) the essentially inviscid instability (described, e.g., in \cite{IM2017}) occurs even in situations where
the classical Couette-Taylor flow is stable; (ii) it can be observed at moderate values of the radial Reynolds number
(though considerably higher than those
for the diverging flow considered in section 3.2): $R\lesssim -105.485$ for $a=2$,  $R\lesssim -80.98$ for $a=4$ and  $R\lesssim -76.203$ for $a=8$.


{\subsection{Critical curves on the $(\Rey_2,\Rey_1)$ plane for diverging flows and weak converging flows}}

\begin{figure}
\begin{center}
\includegraphics*[height=9cm, width=12cm]{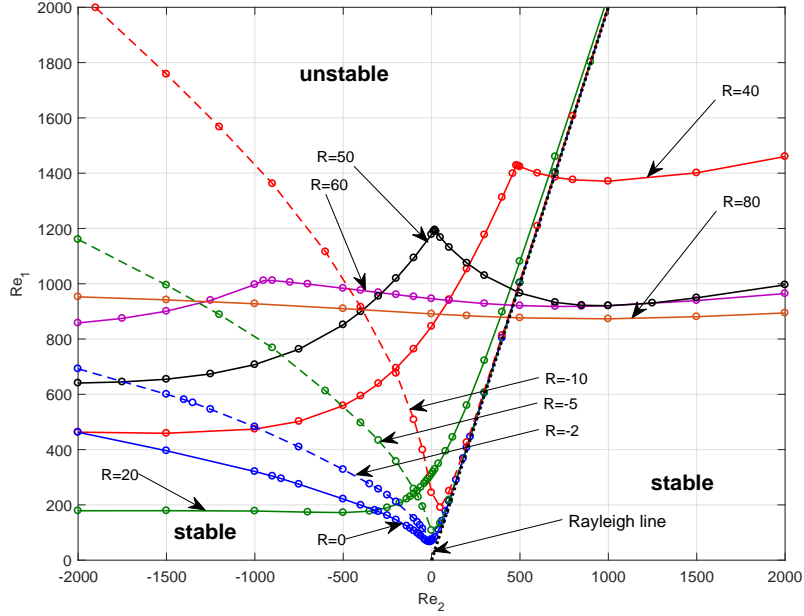}
\end{center}
\caption{Critical curves  on the $(\Rey_2, \Rey_1)$ plane for $a=2$ and several values of $R$.}
\label{IM_fig5}
\end{figure}

\begin{figure}
\begin{center}
\includegraphics*[height=9cm]{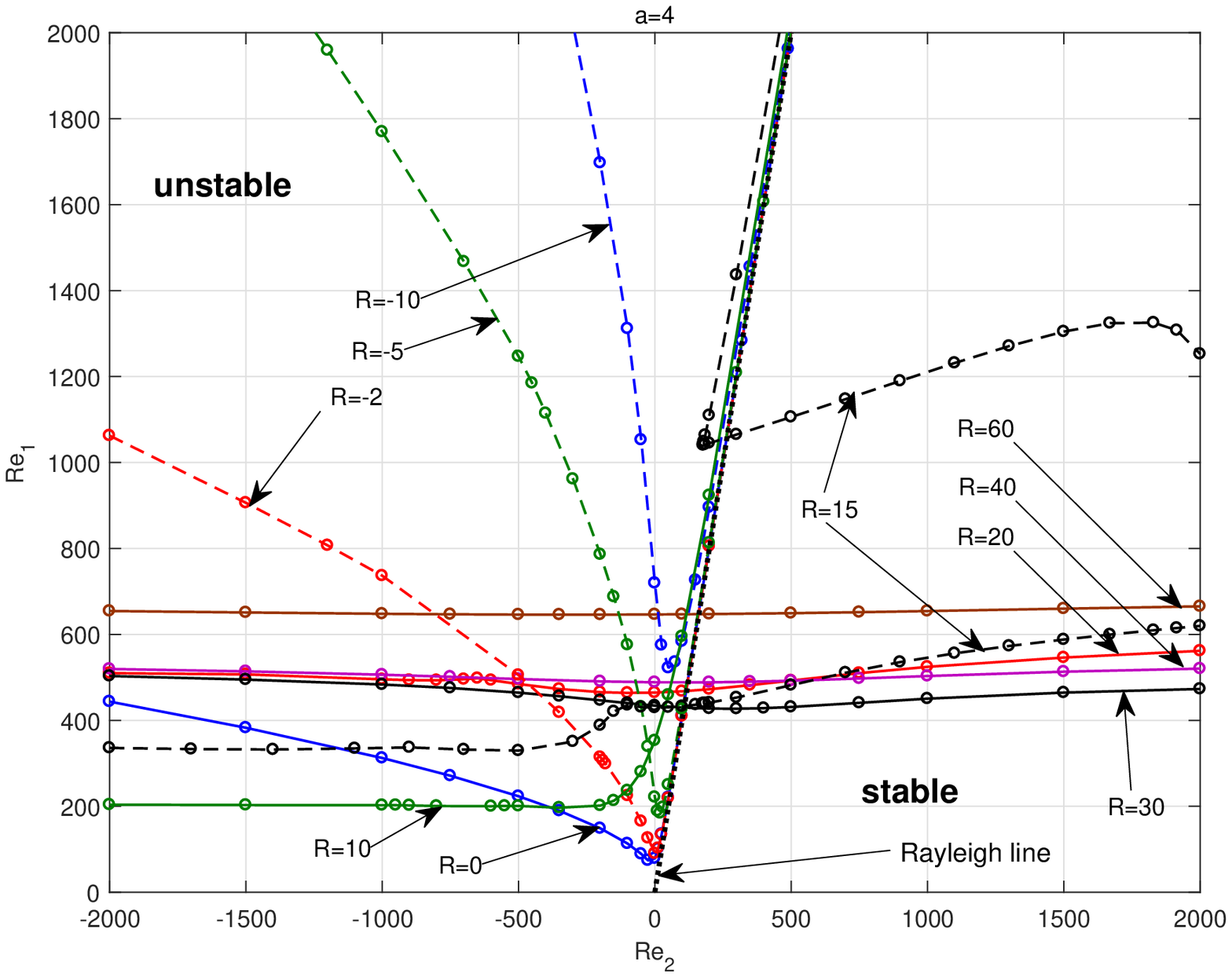}
\end{center}
\caption{Critical curves  on the $(\Rey_2, \Rey_1)$ plane for $a=4$ and several values of $R$.}
\label{IM_fig6}
\end{figure}

\begin{figure}
\begin{center}
\includegraphics*[height=9cm, width=12cm]{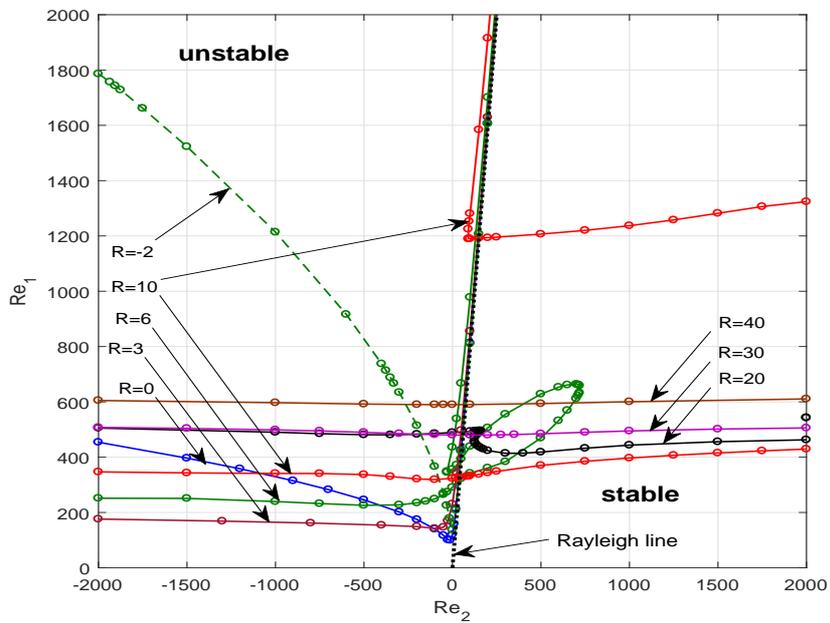}
\end{center}
\caption{Critical curves  on the $(\Rey_2, \Rey_1)$ plane for $a=8$ and several values of $R$.}
\label{IM_fig7}
\end{figure}

Now we consider  critical
curves on the $(\Rey_2,\Rey_1)$ plane.
Since the stability properties of the basic flow do not change under simultaneous change of signs of the angular velocities of the cylinders, it is sufficient to consider only the half-plane $\Rey_1 \geq 0$. The critical curves for several values of $R$,
{which correspond to diverging flows ($R>0$) and weak converging flows ($R<0$ and $\vert R\vert \leq 10$)}, are plotted in Figs. \ref{IM_fig5}-\ref{IM_fig7}
{(strong converging flows will be dealt with separately, in the next section; this is done for the sake of clarity of presentation,
so that the figures are not overcrowded with curves)}.
As before, circles represent the computed points, and solid curves are obtained by linear spline interpolation.
For the classical Couette-Taylor flow, the critical curves separating stability/instability regions on the $(\Rey_2,\Rey_1)$ plane are well-known and agree with the experimental stability  diagram of \citet{Andereck} (see also \cite{Iooss}). The unstable region lies to the left of the Rayleigh line
\[
\Rey_1 = a \, \Rey_2
\]
and the critical curve approaches the Rayleigh line asymptotically as $\Rey_2\to\infty$. It turns out that
the presence of a diverging (or converging) radial flow considerably changes the picture of \citet{Andereck}.

Figure \ref{IM_fig5} shows the critical curves on the $(\Rey_2,\Rey_1)$ plane for $a=2$ and various values of the radial Reynolds number, $R$.
A relatively weak converging radial flow ($R < 0$) does not considerably change the critical curves for positive $\Rey_2$, but, for $\Rey_2 < 0$  (counter-rotating cylinders), it notably shifts the critical curve up (towards higher values of $\Rey_1$).
As the converging radial flow becomes stronger (equivalently, as $\vert R\vert$ increases),
the left part of the critical curve moves further up, so that the instability region rapidly shrinks, while the part of the critical
curve near the
Rayleigh line remains almost unchanged
and lies to the left of the Rayleigh line for all negative values of $R$ shown in Fig. \ref{IM_fig5}. At the same time, the minimum point of the curve, situated
near the vertical axis ($\Rey_2=0$), moves up each time when we increase $\vert R\vert$.
Eventually (for $\vert R\vert\gtrsim 50$), the instability region completely
disappears from Fig. \ref{IM_fig5} (moves up beyond the frame of the figure). So, the converging flow seems to produce
a strong stabilising effect (and this agrees with the earlier results of \cite{Min,Johnson}).

The situation is quite different when a diverging radial flow is imposed. At first, when the radial flow is relatively weak, the only change is that the part of the critical curve corresponding to negative $\Rey_2$ is shifted down (see the critical curve for $R=20$ in Fig. \ref{IM_fig5}). The other part of the curve remains very close to that for
the classical Couette-Taylor flow. As the radial Reynolds number increases further, there comes a point at which the picture radically changes. The critical
curve for $R=40$ in Fig. \ref{IM_fig5} crosses the Rayleigh line, so that the instability region extends to the right of the Rayleigh line where the Rayleigh stability criterion would predict stability.
The reason for such behaviour is the same as in section 3.2: it is a result of the inviscid instability
(see \cite{IM2013a, IM2017, Kerswell}), and the point of non-smoothness of the critical curve (at $\Rey_2\approx 490$) is where the spiral wave inherited from the inviscid instability {becomes more unstable than} the axisymmetric Taylor mode.
An increase in $R$ to $R=50$ results in shrinking of the instability region to the left of the
Rayleigh line and its extension to the right of the Rayleigh line. Further increases in $R$ make the critical curves flatter. For example,
the curve for $R=80$ looks
almost like a horizontal straight line. This means that for sufficiently high radial Reynolds numbers, the stability properties of the basic flow are almost independent
of $\Rey_2$, i.e. the effect of the outer cylinder becomes very weak, which is in agreement with earlier results \cite{IM2013b}.

Critical curves for $a=4$ and $a=8$ shown in Figs. \ref{IM_fig6} and \ref{IM_fig7} look similar to the curves in Fig. \ref{IM_fig5}.
There are a couple of important differences though.
First, the inviscid instability emerges at lower radial Reynolds numbers: this happens for $R\gtrsim 15$ for $a=4$ (Fig. \ref{IM_fig6})
and for $R\gtrsim 6$ for $a=8$ (Fig. \ref{IM_fig7}).
Second, when the instability region crosses Rayleigh line, it covers a tongue-like area to the right of it
(see the critical curve
for $R=6$ in Fig. \ref{IM_fig7}). It grows when $R$ increases (see the curves for $R=10$ in Fig. \ref{IM_fig7} and for $R=15$ in Fig. \ref{IM_fig6}), and
its upper and right boundaries very quickly move beyond the upper and right edges of Figs. \ref{IM_fig6} and \ref{IM_fig7} ($R\gtrsim 20$ in Fig.
\ref{IM_fig6} and $R\gtrsim 15$ in Fig. \ref{IM_fig7}). Again, for sufficiently high $R$, the critical curves look like horizontal straight lines.
\begin{figure}
\begin{center}
\includegraphics*[height=9cm, width=12cm]{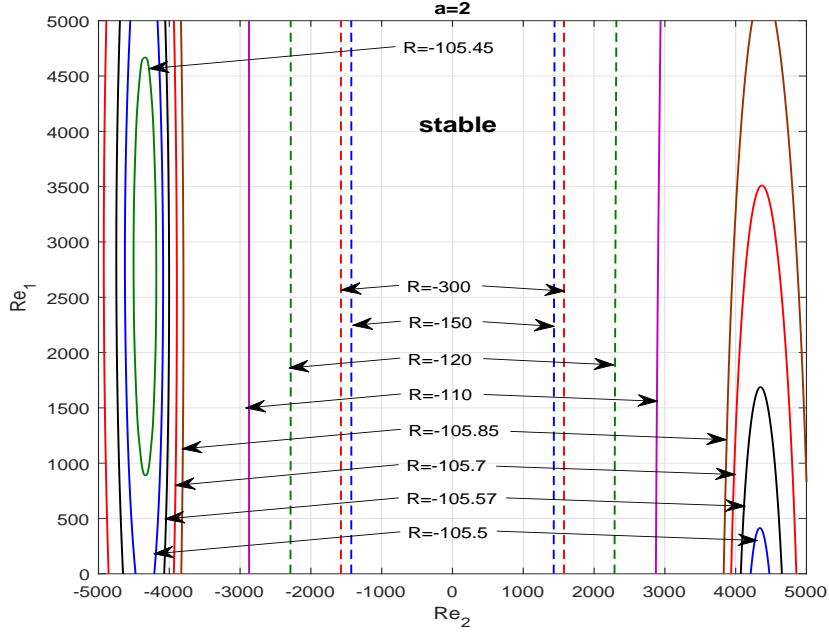}
\end{center}
\caption{Stability/instability regions on the $(\Rey_2, \Rey_1)$ plane for $a=2$ and several negative values of $R$.}
\label{IM_fig8}
\end{figure}\begin{figure}
\begin{center}
\includegraphics*[height=9cm, width=12cm]{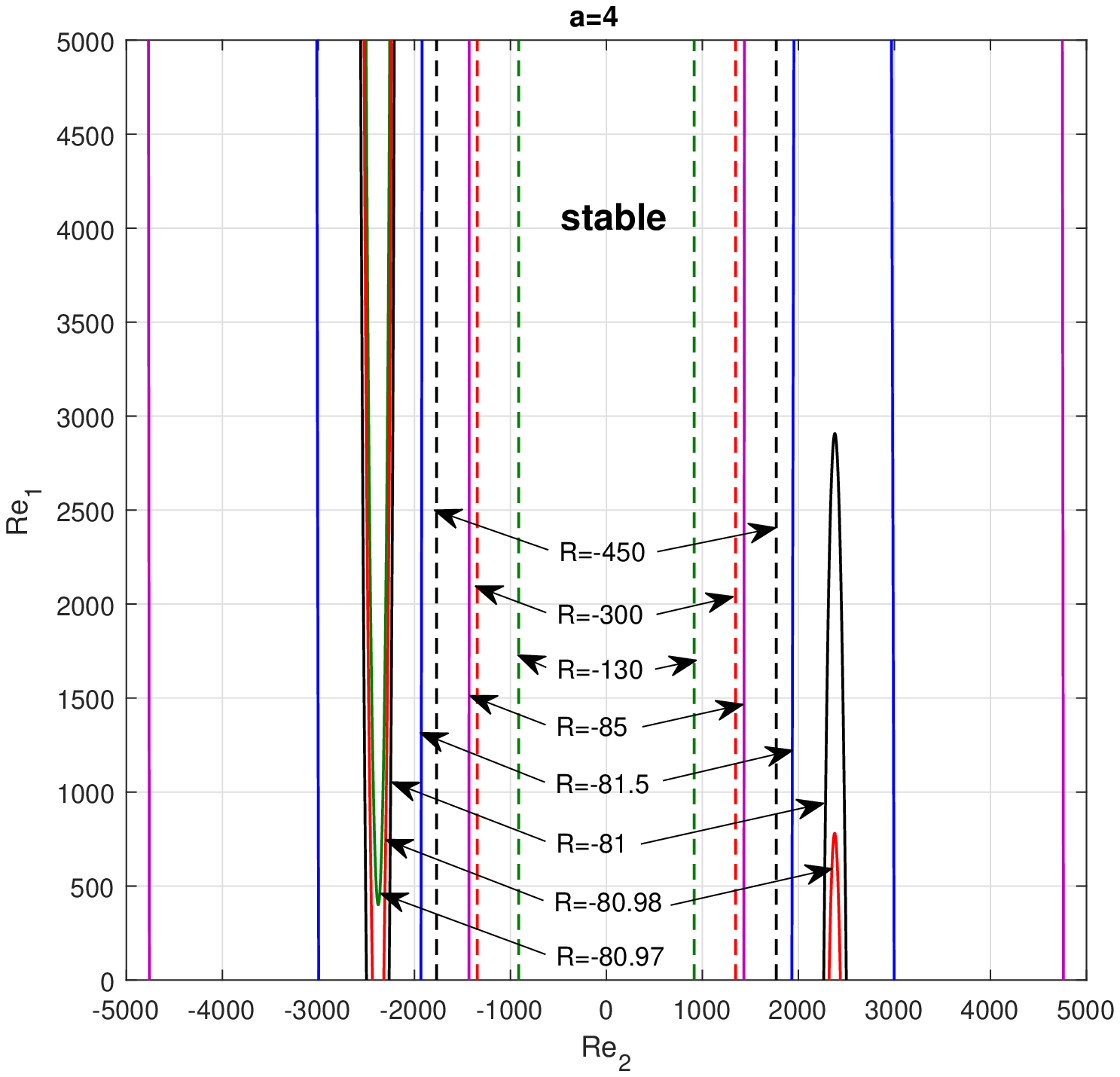}
\end{center}
\caption{Stability/instability regions on the $(\Rey_2, \Rey_1)$ plane for $a=4$ and several negative values of $R$.}
\label{IM_fig9}
\end{figure}\begin{figure}
\begin{center}
\includegraphics*[height=9cm, width=12cm]{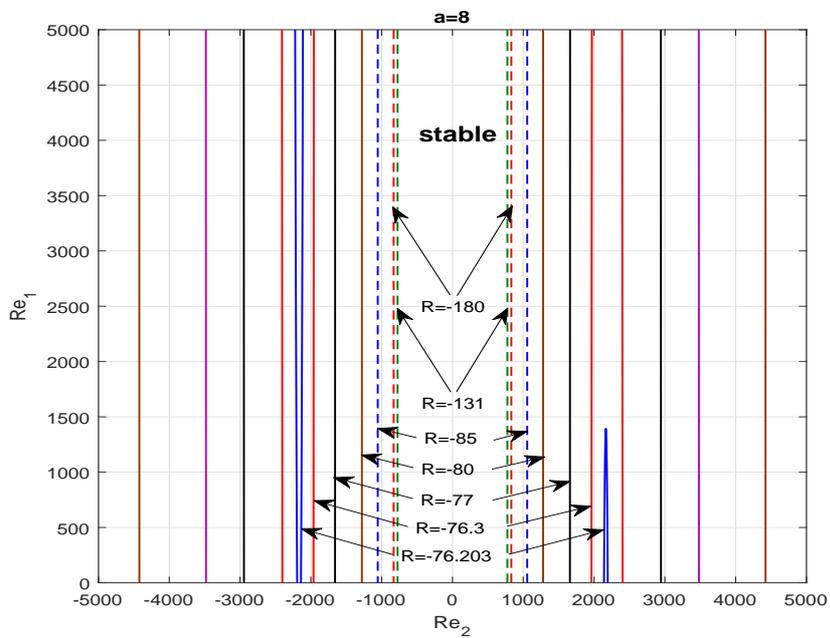}
\end{center}
\caption{Stability/instability regions on the $(\Rey_2, \Rey_1)$ plane for $a=8$ and several negative values of $R$.}
\label{IM_fig10}
\end{figure}

{It is evident from Figs. \ref{IM_fig5}-\ref{IM_fig7} that for the cylinders rotating in opposite directions
($\Rey_2 < 0$), a weak diverging radial flow is destabilizing: the critical curves for (relatively) small positive $R$ are
below the curve for the classical Couette-Taylor flow provided that the magnitude of $\Rey_2$ is sufficiently large
($\Rey_2 \lesssim -100$). Figures \ref{IM_fig5}-\ref{IM_fig7} also show a sufficiently strong diverging flow has
stabilizing effect. Therefore, there must be an optimal value of $R$ corresponding to the lowest critical value of $\Rey_1$ for each
value of $\Rey_2 < 0$. We have computed $\Rey_{1*}=\Rey_{1cr}(R_*)=\min_{R\geq 0}\Rey_{1cr}(R)$ for several values
of $\Rey_2$ for $a=4$, as well as
the difference, $\Delta \Rey_1$, between the critical value of $\Rey_1$ for the classical Couette-Taylor flow, $\Rey_{1,0}$,
and $\Rey_{1*}$ (i.e.
$\Delta \Rey_1$ defined as
$\Delta \Rey_1=\Rey_{1,0}- \Rey_{1*} \, $). The results are shown
in Table \ref{tab1}. }
\begin{table}
  \begin{center}
{
  \begin{tabular}{l|l|l|l|l|l|l|l}
$\Rey_2$ &-100 &-200 &-400 &-600 &-800 &-1000 &-1400 \\
\hline
$R_*$    &2.47 &2.89 &3.48 &3.82 &4.04 &4.20 &4.52\\
\hline
$\Rey_{1*}$ &81.6 &91.8 &102 &108 &113 &117 &122\\
\hline
$\Delta\Rey_{1}$ &31.6 &56.9 &99.5 &135 &167 &196 &247
\end{tabular}
}
  \caption{{Values of $R_*$, $\Rey_{1*}$ and
  $\Delta \Rey_1$ for several negative values
of $\Rey_2$ for $a=4$.}}
  \label{tab1}
  \end{center}
\end{table}
{Evidently, the minimal critical value of $\Rey_1$, the optimal value of $R$ and the difference,
$\Delta \Rey_{1}$, all grow when $\Rey_2$ decreases. However, the growth in $R_*$ and $\Rey_{1*}$ is pretty slow, while the gap between
$\Rey_{1*}$ and the critical $\Rey_1$ for the classical Couette-Taylor flow widens much faster
(it increases by a factor of $7.8$ as $\Rey_2$ varies from $-100$ to $-1400$. So, the destabilizing effect of
the weak diverging radial flow is pretty strong and it becomes much stronger when $\Rey_2$ decreases. }


{\subsection{Critical curves on the $(\Rey_2,\Rey_1)$ plane for strong converging flows}}

{One of the findings of the preceding section is the stabilizing effect of the converging radial flow: the
instability domain on the $(\Rey_2,\Rey_1)$ plane rapidly shrinks when $R$ decreases from $0$. }
However, we also saw in section 3.4 that the flow with a fixed inner cylinder ($\Rey_1 =0$) becomes unstable for a sufficiently strong converging radial flow due to the inviscid instability \cite{IM2013a}. So, unstable regions should also appear on the
$(\Rey_2,\Rey_1)$ plane provided that $R<0$ and $\vert R\vert$ is sufficiently high. The results of calculations are presented in Figs. \ref{IM_fig8}-\ref{IM_fig10}. Figure \ref{IM_fig8}
shows the instability regions for $a=2$. {When the converging flow is sufficiently strong ($R\approx -105.422$), a vertically elongated oval instability region, which is at first very small, appears in the second quadrant of the plane. As $R$ decreases, this oval region rapidly grows. The instability region for $R= -105.45$ is shown in Fig. \ref{IM_fig8}. As $R$ decreases further, the upper and lower parts of the oval instability region move beyond the edges of the figure (see the curve for
$R=-105.5$ in Fig. \ref{IM_fig8}).} At the same time the upper part of another oval instability region emerges in the first quadrant.
(It should be noted here
that the picture on the upper half-plane can be continued to the lower half-plane using the central symmetry with respect to the origin. This means that two oval regions appear simultaneously in the second and forth quadrants of the whole $(\Rey_2,\Rey_1)$ plane
and then grow in size penetrating into the first
and third quadrants when $R$ decreases.) {As the strength of the converging flow increases further},
the ovals continue to grow in size and their centers slowly drift away from the vertical
axis ($\Rey_2=0$). As a result the region of stability near the vertical axis, while shrinking at first, never completely disappears, and starts growing in width
at $R\approx -170$. For all curves in Fig. \ref{IM_fig8}, the critical modes are two-dimensional ($k_{cr}=0$). On each curve,
the critical azimuthal wave number does not change (so that the critical curves appear to be smooth), and $n_{cr}=2$ for all curves except for
$R=-300$, for which $n_{cr}=3$. Critical curves for $a=4$ and $a=8$ are presented in Figs. \ref{IM_fig9} and \ref{IM_fig10}. They look similar to those in
Fig. \ref{IM_fig8}. For all critical curves in Figs. \ref{IM_fig9} and \ref{IM_fig10}, $k_{cr}=0$. In Fig. \ref{IM_fig9}, $n_{cr}=1$ for all
$R$ except $R=-300$ and $R=-450$, for which $n_{cr}=2$. In Fig. \ref{IM_fig10}, $n_{cr}=1$ for all curves.
It is evident from  Figs. \ref{IM_fig9} and \ref{IM_fig10} that, compared with the case of $a=2$, the instability occurs
for slightly weaker radial flows (at $R\approx -80.97$ for $a=4$ and
$R\approx -70.203$ for $a=8$) and that the oval instability regions become more elongated for larger $a$ and lie closer to
the vertical axis $\Rey_2=0$. These observations lead us to a conclusion that, although the effect of a weak converging flows
is stabilising, it becomes destabilising provided that the converging radial flow is sufficiently strong.

\vskip 3mm

\section{Discussion}


We have investigated the linear stability of steady viscous incompressible flows between rotating porous cylinders with a radial flow for a wide range of
the key parameters of the flow: the ratio of the radii of the cylinders, $a$, the radial Reynolds number, $R$, and the inner and outer azimuthal Reynolds numbers,
$\Rey_1$ and $\Rey_2$. This resulted in a detailed picture of the effect of a radial flow.
In particular, we have obtained critical curves on the $(\Rey_2,\Rey_1)$ plane for $a=2, 4, \ \hbox{and} \ 8$ and various values of $R$. These critical curves
show that a relatively weak diverging radial flow ($R>0$) can have both stabilising and destabilising effects depending on $\Rey_2$. The critical curves for $R=20$ in
Fig. \ref{IM_fig5}, for $R=10$ in Fig. \ref{IM_fig6} and for $R=3$ in \ref{IM_fig7} show that for negative $\Rey_2$ (counter-rotating cylinders), the basic flow is more stable (than the classical Couette-Taylor flow) for small to moderate negative $\Rey_2$, but it becomes more unstable
for large negative $\Rey_2$; for positive $\Rey_2$
(co-rotating cylinders), the critical curves remain close to the classical Couette-Taylor curves.
As the radial Reynolds number increases, the critical modes take the form of spiral or planar azimuthal waves rather than the classical
Taylor vortices. These waves are what is left of the inviscid instability (studied earlier in \cite{IM2013a, Kerswell, IM2017}) when
the viscosity is taken into account.
The key consequence of this inviscid instability is that some
flows in the region to the right of the Rayleigh line (which are stable in the classical Couette-Taylor flow) become unstable if a sufficiently strong
diverging radial flow is present. The most interesting  (and unexpected) result here is that the inviscid instability survives for radial Reynolds numbers as low as
$R= 40$ for $a=2$ (see Fig. \ref{IM_fig5}), $R=15$ for $a=4$ (see Fig. \ref{IM_fig6}) and $R=6$ for $a=8$ (see Fig. \ref{IM_fig7}).
This makes an experimental observation of the inviscid instability feasible.

In the case of converging radial flow(see Figs. \ref{IM_fig4}, \ref{IM_fig8}-\ref{IM_fig10}), the inviscid instability again invades the regions where
the classical Couette-Taylor flow is stable (this occurs, e.g., when the inner cylinder is not rotating). However, the converging radial flow has
to be rather strong to produce the instability.
We found that it can be observed at $R\lesssim -105.45$ for $a=2$ (Fig. \ref{IM_fig8}), $R\lesssim -80.97$ for $a=4$ (Fig. \ref{IM_fig9}) and $R\lesssim -76.203$ for $a=8$ (Fig. \ref{IM_fig10}). The absolute values of these numbers are considerably higher than their counterparts for the diverging flows, but still moderate, so that the inviscid instability in converging flows can, hopefully, be observed in experiments.

Within the range of parameters considered in this paper, the picture of linear instabilities that emerges from Figs. \ref{IM_fig5}-\ref{IM_fig10} is complete
in the sense that there are no other linear instabilities. However, as was discussed in \cite{IM2013b},
the viscous boundary layers that appear near the outflow part of the boundary
can also be unstable. In particular,
it has been shown there that, at high radial Reynolds numbers, this instability of the boundary layer is equivalent to the instability of
the asymptotic suction profile (see, e.g., \cite{Hughes,Hocking,Doering2000}).
This viscous boundary layer instability is, however, well separated from all the instabilities discussed in the present paper:
the former requires very large values of $\vert \gamma_1-\gamma_2\vert$ ($\vert \gamma_1-\gamma_2\vert > 5 \cdot 10^4$), while
the latter occur at moderate values of $\gamma_1$ and $\gamma_2$.

There are still quite a few open problems in this area. For example, it would be interesting to develop a weakly nonlinear theory of the oscillatory `inviscid'
instabilities considered in the present paper. The recent theory of \citet{Martinand2017} (where the effects of both radial and axial flows were considered)
is not directly applicable to the case without axial flow. This is because the axial flow breaks the mirror symmetry of the problem, so that
the eigenvalues associated with neutral modes are simple, while they are double when there is no axial flow. The weakly nonlinear analysis taking
into account this multiplicity and the symmetry breaking by a weak axial
flow is a topic of a continuing investigation.
Another interesting question is whether the instabilities discussed here
have any relevance to accretion disks in astrophysics. The weakly compressible analysis of \citet{Kerswell} is not sufficient (as was pointed out by Kerswell himself) because the azimuthal velocity in thin accretion disks is believed to be highly supersonic (see, e.g., \cite{Frank}). This is also a problem for a future investigation.



\end{document}